\documentclass[aps,prc,twocolumn,groupedaddress,showpacs]{revtex4}

\usepackage{graphicx}
\usepackage{latexsym}
\usepackage{bm}
\usepackage{amsmath}

\newcommand{\comment}[1]{}

\begin{document}


\title{Energy loss in high energy heavy ion collisions from the Hydro+Jet model}

\author{Tetsufumi Hirano}
\affiliation{Physics Department, University of Tokyo,
  Tokyo 113-0033, Japan}
\author{Yasushi Nara}
\affiliation{%
 RIKEN BNL Research Center, Brookhaven National Laboratory,
                Upton, N.Y. 11973, U.S.A.
}

\date{\today}

\begin{abstract}
We investigate the effect of energy loss of jets 
in high energy heavy ion collisions
by using a full three-dimensional space-time evolution of a fluid
combined with (mini-)jets that are explicitly evolved
in space-time.
In order to fit the $\pi^0$ data for the Au+Au collisions
 at $\sqrt{s_{NN}}=130$ GeV,
the space-time averaged energy loss
 $dE/dx(\tau \leq \text{3 fm/}c)=0.36$ GeV/fm
 is extracted within the model.
It is found that most energy loss occurs
at the very early time less than 2 fm/$c$
 in the QGP phase
and that energy loss in the mixed phase is negligible
within our parameterization for jet energy loss.
This is a consequence of strong expansion of the system.

\end{abstract}


\maketitle


Measurements of high $p_{\mathrm{T}}$ hadrons
 at the Relativistic Heavy Ion Collider (RHIC)
and Large Hadron Collider (LHC)
 may provide insight into the excited partonic matter, often
called a quark gluon plasma (QGP), produced in heavy ion
collisions~\cite{QMproc}.
Because jets have to traverse the excited matter, their spectra
should be changed compared to the elementary hadron-hadron data.
This energy loss of jets has been proposed as a possible
signature of the QGP phase
~\cite{Gyulassy:1990ye}.
Over the past year, a lot of work have been devoted to study
the propagation of jets through
 QCD matter~\cite{Wiedemann,Zakharov,Baier,Levai}.

Recently, hadronic transverse momentum distribution
in Au + Au collisions
at $\sqrt{s_{NN}}=130$ GeV has been measured at RHIC and
found that these spectra show the depletion
 at high transverse momentum~\cite{phenix_pi0}.
Comparison of the data with
pQCD parton model based calculations~\cite{Wang:2001gv,Fai,Levai2}
and a transport theoretical study~\cite{Nara:2001zu}
 shows the indication of energy loss
in central Au + Au collisions at RHIC energies.
Phenomenological studies~\cite{Wang:2001gv,jamal}
 suggest that the space-time averaged energy loss
yields small values of  $dE/dx \sim 0.25$--$0.3$ GeV/fm.
In addition,
jet quenching is found to be the main contribution
on the elliptic flow parameter in the large transverse momentum
region~\cite{v2jet1,v2jet2}.
Jet energy loss is modeled by modifying
the fragmentation functions in pQCD parton model and
usually neglecting the dynamical effects.

It might be important to consider dynamical effects on the jet interactions.
The effects of the expansion of the system on the energy loss
and the azimuthal asymmetry
have been studied~\cite{v2jet1,GVWH,EWANG}.
Hydrodynamics provides the space-time evolution of
thermalized partonic/hadronic matter produced
 in heavy-ion collisions~\cite{KHHH,Teaney:2000cw,Hirano:2001eu,Hirano:2001yi}.
Recently, one of the authors (T.H.) developed a fully three-dimensional
hydrodynamic model in Cartesian coordinate~\cite{Hirano:2000eu} and also in Bjorken coordinate~\cite{Hirano:2001eu} and, moreover, took into account
the picture of the early chemical freeze-out in this
model~\cite{Hirano:2002ds}.
Hydrodynamic model calculations at the RHIC energies suggest that
the lifetime of the  pure QGP phase is to be about 4 fm/$c$
and that the mixed phase exists up to 9--10 fm/$c$~\cite{Hirano:2002ds}.
The density of the system drops rapidly due to the longitudinal
expansion. 

In this Letter,
  we explore the dynamical effects on
 the energy loss of jets by taking into account
the full 3D space-time evolution of a fluid.
The hydrodynamic model is combined with the jets (Hydro + Jet model)
 which
are calculated from the pQCD parton model
and explicitly propagated in space-time with fluid elements.


We use full 3D hydrodynamics~\cite{Hirano:2001eu}
 in which initial parameters are fixed for Au+Au
collisions at $\sqrt{s_{NN}}=130$ GeV~\cite{Hirano:2002ds}.
The transverse profile of the initial energy density is assumed to
scale with the number of binary collisions \cite{KHHH}.
For 5\% central collisions, we choose the impact parameter as $b=2.4$ fm and the maximum initial energy density at the initial time $\tau_0=0.6$ fm/$c$ as 33.7 GeV/fm$^3$.
These parameters lead us to reproduce transverse momentum
spectra of charged hadrons \cite{Adler:2001yq} up to 1.0--1.5 GeV/$c$.
It is found that the $p_{\mathrm{T}}$ slope is insensitive to the thermal
freeze-out temperature $T^{\mathrm{th}}$ when we take
into account the early chemical freeze-out~\cite{Hirano:2002ds}.
We fix $T^{\mathrm{th}} = 140$ MeV throughout this Letter.

We include hard partons using pQCD parton model,
\begin{equation}
  {d \sigma_{\mathrm{jet}}\over d p^2_\mathrm{T}dY_1dY_2}
  = K\sum_{a,b} x_1x_2f_a(x_1,Q^2)f_b(x_2,Q^2)
    {d\sigma_{ab} \over d\hat{t}},
\end{equation}
where $Y_1$ and $Y_2$ are the rapidities of the scattered partons
and $x_1$ and $x_2$ are the fractions of momentum of the initial partons.
The parton distribution functions $f_a(x,Q^2)$ are taken to be CTEQ5 leading
order~\cite{cteq5}.
We use $Q^2= p_{\mathrm{T}}^2$ for the evaluation of parton distribution.
The minimum momentum transfer $p_{\text{T,min}} = 2.0$ GeV/$c$ is assumed.
The summation runs over all parton species and
relevant leading order QCD processes
\begin{eqnarray}
   q + q' &\to& q + q', \quad q + \bar{q} \to q' + \bar{q}',\\
   q + \bar{q} &\to& g + g \quad q + g \to q + g,\\
   g + g &\to& q + \bar{q}, \quad g + g \to g + g.
\end{eqnarray}
are included
in addition with the initial and final state radiation
to simulate the emission of multiple soft gluons.
Gaussian primordial transverse momentum  $k_{\mathrm{T}}$
 distribution with the width of
 $\langle k_{\mathrm{T}}^2\rangle = 1$ GeV$^2$
 is assigned to the shower initiator
 in the QCD hard $2\to2$ processes.
We use PYTHIA 6.2~\cite{pythia} to simulate each hard scattering
 in the actual calculation.
A factor $K$ is used for the higher order corrections.
We chose $K=2$ to fit the UA1 data of $p\bar{p}$ at $\sqrt{s}=200$
GeV~\cite{ua1}.
In order to convert hard partons into hadrons, we use an independent fragmentation
model using PYTHIA after hydro simulations.
We have checked that this hadronization model provides good agreement
with the transverse spectra of charged hadrons above
 $p_{\mathrm{T}} = 1$ GeV/$c$ in $p\bar{p}$ data~\cite{ua1}.

The number of hard partons is assumed to scale with the number
of hard scattering which is estimated by using Woods-Saxon nuclear density.
This assumption is consistent with the peripheral 
Au+Au collision at $\sqrt{s_{NN}}=130$ GeV~\cite{Fai}.
The space coordinate of a parton in longitudinal direction is
 taken to be $\eta_{\mathrm{s}} = Y$,
 where $\eta_{\mathrm{s}}=(1/2)\log[(t+z)/(t-z)]$,
and $Y$ is a momentum rapidity.
The transverse coordinate of a parton is specified by the
number of binary collision distribution for two Woods-Saxon distributions.
Most hard partons are produced
with the formation time of $\sim 1/p_{\mathrm{T}}$
assuming the uncertainty relation.
Since the initial time of the hydrodynamical simulation is $\tau_0=0.6$ fm/$c$,
partons are assumed to travel freely up to $\tau_0$.
We neglect the nuclear shadowing effect in the present work for simplicity,
because it is small
 at high transverse momentum~\cite{Wang:2001gv,jamal}.

\begin{figure}[t]
\includegraphics[width=3.3in]{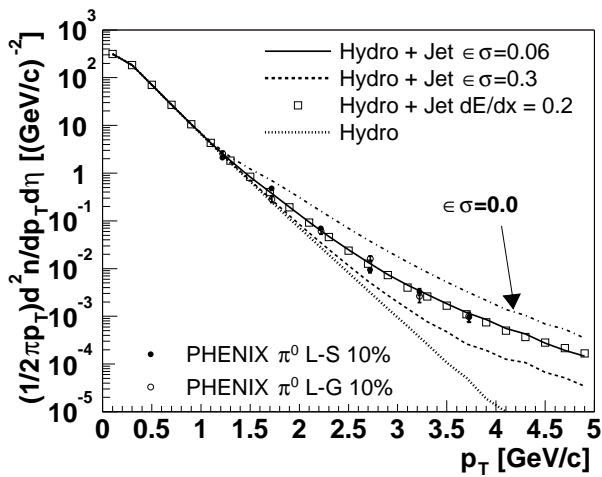}
\caption{Comparison of $\pi^0$ spectra
from hydro + jet
with the energy loss parameter $\epsilon\sigma=0.06$ GeV fm$^2$ (solid line),
hydro + jet with $\epsilon\sigma=0.3$ GeV fm$^2$ (dashed line), 
and hydro only (dotted line)
for $b=3.35$ fm
to the PHENIX data~\cite{phenix_pi0} for centrality of 10\%.
The result of hydro + jet for $dE/dx=0.2$ GeV/fm calculation is shown
in open squares.
}
\label{fig:dndptpi0_central}
\end{figure}
\begin{figure}
\includegraphics[width=3.3in]{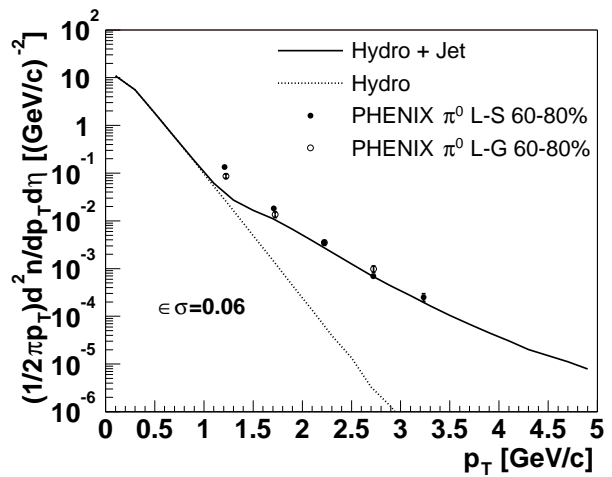}
\caption{Comparison of $\pi^0$ spectra
from hydro + jet (solid line) and hydro only (dotted line)
with the energy loss parameter $\epsilon\sigma=0.06$ GeV fm$^2$
and $b=12.1$ fm
to the PHENIX data~\cite{phenix_pi0} for peripheral collision (60-80\%).
}
\label{fig:dndptpi0_peripheral}
\end{figure}
%
%
We assume that the form of energy loss is simply
\begin{equation}
 {dE \over dx} = \epsilon \sigma \rho(\tau,\bm{r}),
\label{eq:energyloss}
\end{equation}
where $\epsilon$ is an energy loss per scattering and
$\sigma$ is a parton-parton cross section.
From a hydrodynamic simulation, we obtain the space-time evolution
of temperature $T(\tau,\bm{r})$.
A parton density $\rho(\tau,\bm{r})$ in the QGP phase
 is calculated from $T(\tau,\bm{r})$. 
For the mixed phase at $T=T_\mathrm{c}$(=170 MeV),
 we put $\rho(\tau,\bm{r})= f_{\mathrm{QGP}}(\tau,\bm{r})\rho(T_\mathrm{c})$
 where the fraction of the QGP phase
 in the mixed phase $f_{\mathrm{QGP}}(\tau,\bm{r})$ is calculated
 from energy density $\varepsilon(\tau,\bm{r})$
 and the maximum and minimum energy densities
 in the mixed phase 
$\varepsilon_{\mathrm{QGP}}$
 and $\varepsilon_{\mathrm{had}}$ as
\begin{equation}
f_{\mathrm{QGP}}(\tau,\bm{r})
  = {\varepsilon(\tau,\bm{r})-\varepsilon_{\mathrm{had}} \over
     \varepsilon_{\mathrm{QGP}}-\varepsilon_{\mathrm{had}}}.
\end{equation}
We assume that the energy loss for a quark jet is
 half of the energy loss of a gluon~\cite{wang2}.
In the present work, we neglect the possible space-time variation
of parton-parton cross section and regard the 
product $\epsilon\sigma$
as an adjustable free parameter in the model.
Feedback of the energy to fluid elements is ignored, because we
use
the relatively small values of energy loss parameter used in the work.


In Fig.~\ref{fig:dndptpi0_central},
 the transverse momentum distributions
of neutral pions for central 10\% Au + Au collisions at $\sqrt{s_{NN}}=130$
GeV from PHENIX experiment~\cite{phenix_pi0}
are compared to the results of the hydro + jet model
with different energy loss parameters $\epsilon\sigma=$ 0, 0.06 and 0.3
GeV fm$^2$
 together with the
hydro result by dotted line.
In the calculation, we choose the impact parameter
$b=3.35$ fm for 0--10\% central events. The corresponding number
of binary collisions is $N_{\mathrm{binary}}=906$.
In order to connect the contribution from jets and a fluid smoothly,
low $p_{\mathrm{T}}$ pions from jets are cut by
 a switch function
$\{1+\tanh[3(p_{\mathrm{T}}-1.5 \mathrm{GeV})]\}/2$~\cite{v2jet1}.
We reproduce $p_{\mathrm{T}}$ spectrum for $\pi^0$
 by choosing $\epsilon\sigma = 0.06$ GeV fm$^2$,
while the result which is not taken into account a energy loss
overestimates the neutral pion spectrum.
%
The hydro + jet result for $dE/dx=0.2$ GeV/fm
which is obtained by neglecting the density dependence
in Eq.~(\ref{eq:energyloss})
is also shown in Fig.~\ref{fig:dndptpi0_central}.
We note that the hydro + jet result for $\epsilon\sigma=0.3$ GeV fm$^2$
is reproduced by the constant energy loss of $dE/dx=1.0$ GeV/fm.
As shown in Fig.~\ref{fig:dndptpi0_peripheral},
$\pi^0$ spectra for peripheral (60-80\%) Au + Au collisions~\cite{phenix_pi0}
can be reproduced 
only by changing the impact parameter to $b=12.1$ fm ($N_{\mathrm{binary}}=20$)
and leaving the other parameters.
The hydro + jet result in peripheral collisions
does not change when energy loss is not
included, because the parton density and the volume of the QGP phase are small.
On the other hand, the hydro + jet model overpredicts the data
without energy loss in the central collisions.
The deviation from data in the transverse momentum range below 1.5 GeV/$c$
in the peripheral collisions
might be reasonable, because we do not expect that local thermalization
is achieved in such a peripheral collision and that hydrodynamics
is reliable.

\begin{figure}
\includegraphics[width=3.0in]{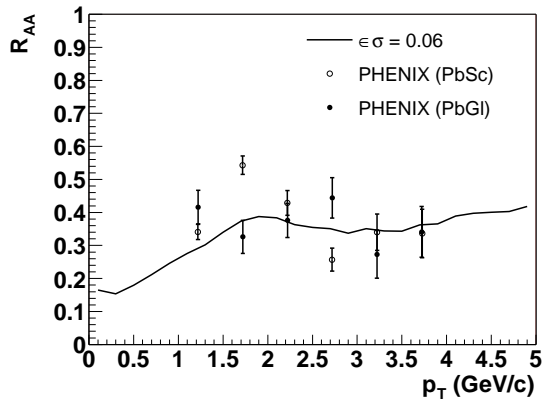}
\caption{The ratio $R_{AA}$ for neutral pions in Au+Au collisions with $b=3.35$ fm. The error bars of the PHENIX data \cite{phenix_pi0} contain the statistical error only.
}
\label{fig:RAA}
\end{figure}

The jet quenching is quantified by the ratio of the particle
yield in A+A collisions to the one in p+p collisions scaled up
by the number of binary collisions
\begin{equation}
R_{AA} = \frac{d^2N^{A+A}/dp_T d\eta}{N_{\mathrm{binary}}
d^2N^{p+p}/dp_T d\eta}.
\end{equation}
The transverse momentum dependence of $R_{AA}$ can provide information
about the mechanism of jet quenching \cite{jeon}.
In Fig.~\ref{fig:RAA}, the result from the Hydro+Jet model with
$\epsilon\sigma = 0.06$ GeV fm$^2$ is compared to the PHENIX data
\cite{phenix_pi0}.
Our result is almost flat in the range $2 < p_T < 4$ GeV/$c$.

\begin{figure}
\includegraphics[width=3.3in]{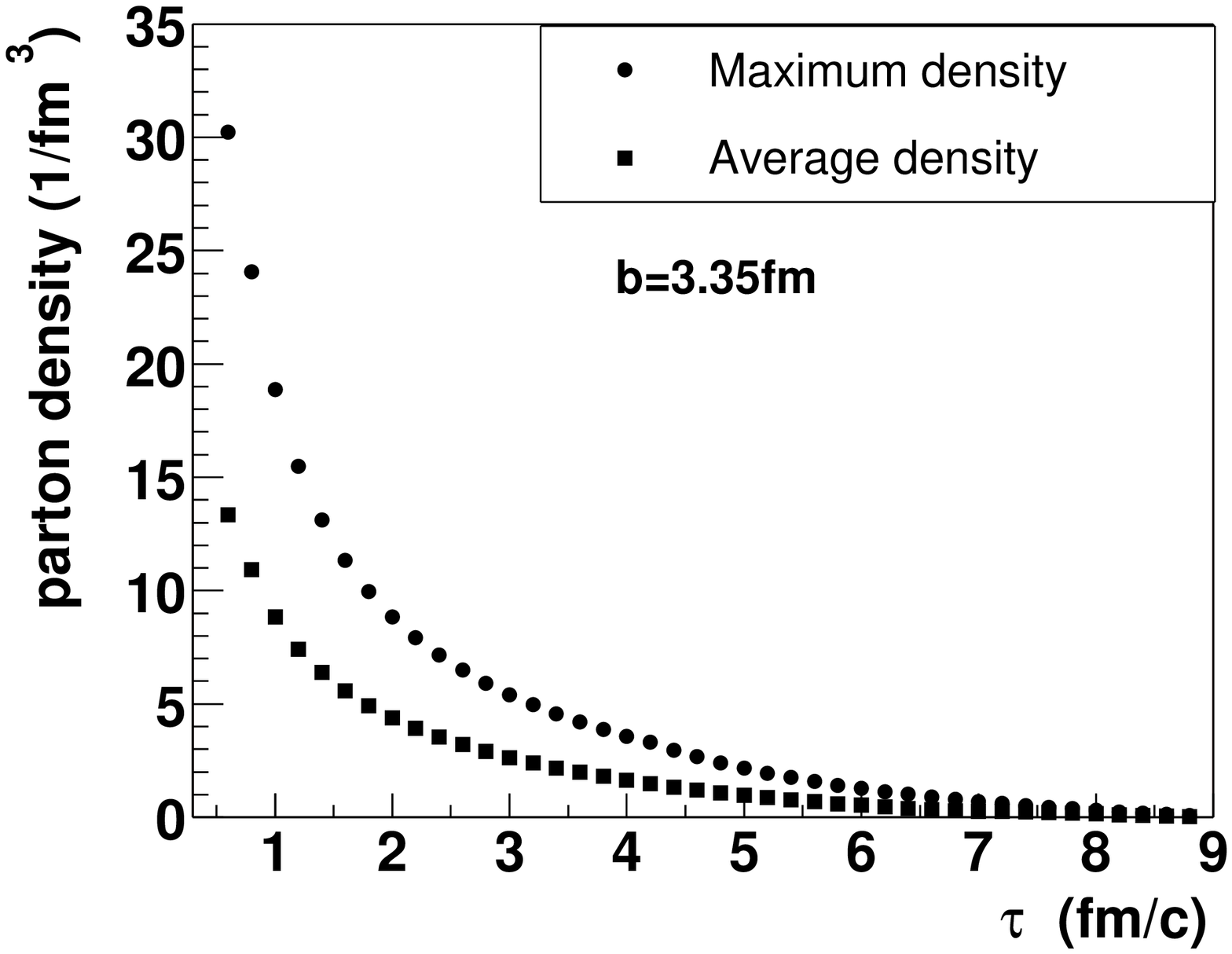}
\includegraphics[width=3.3in]{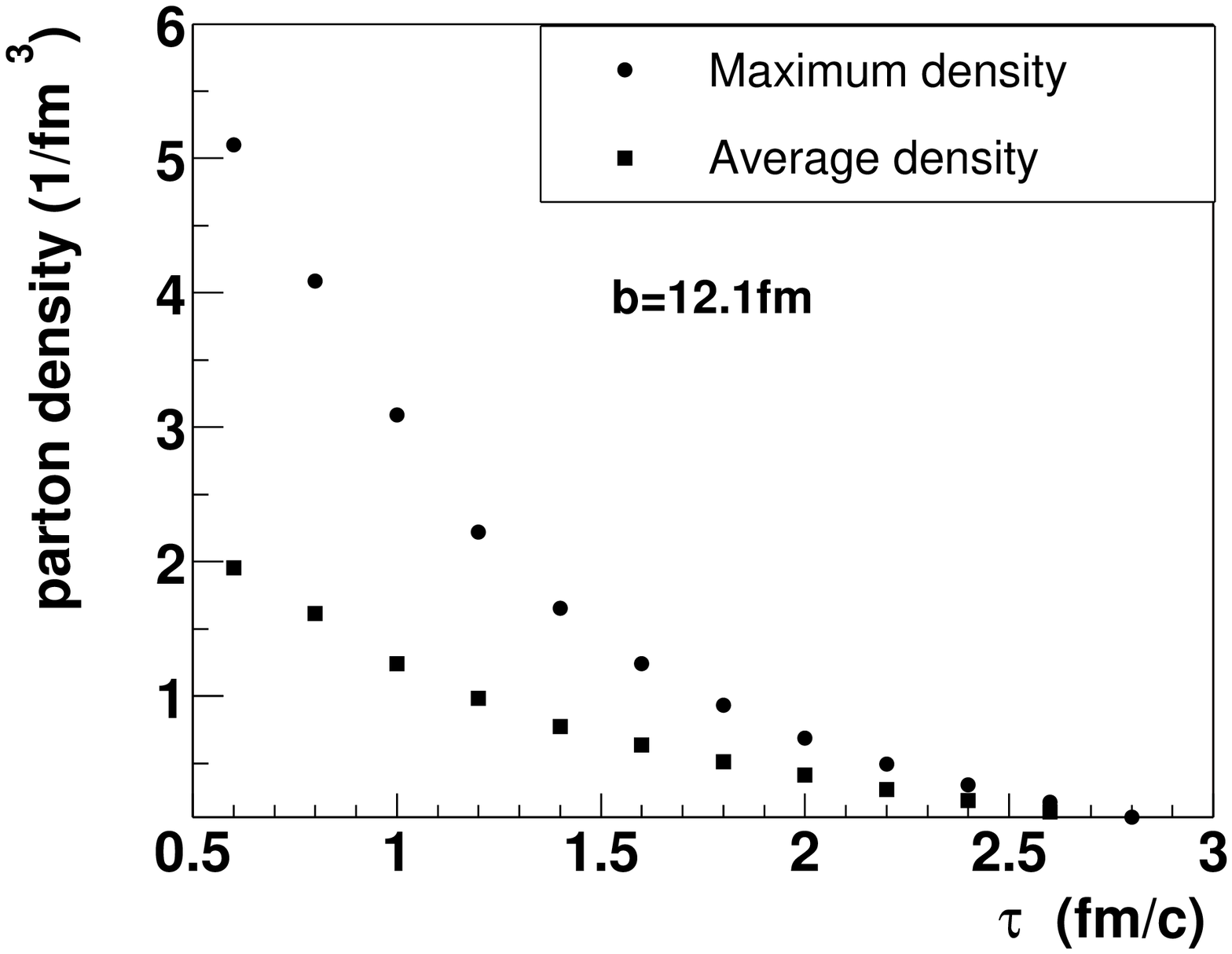}
\caption{Maximum (circles) and average parton density (squares)
 as a function of proper time at impact parameter $b=3.35$ fm (upper)
 and $b=12.1$ fm (lower).}
\label{fig:timeevoldens}
\end{figure}

\begin{figure}
\includegraphics[width=3.0in]{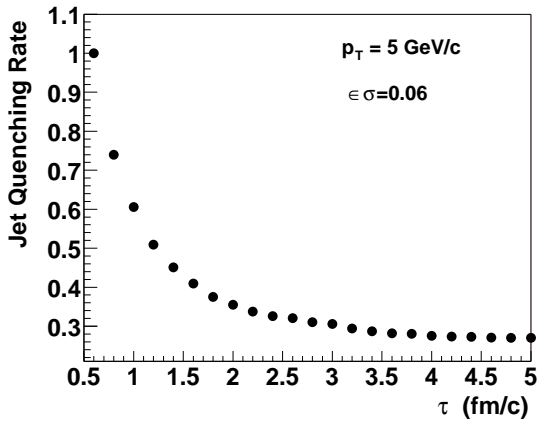}
\caption{Jet quenching rate
defined as the ratio of the numbers of jets $N(\tau)/N(\tau_0)$
at $p_{\mathrm{T}} = 5$ GeV/$c$
 as a function of proper time for $b=3.35$ fm.
The energy loss parameter $\epsilon\sigma=0.06$ GeV fm$^2$
is used.}
\label{fig:jetquenchingrate}
\end{figure}

Let us now turn to the study of the dynamical effects on the jet energy loss.
The maximum and the average thermalized parton densities
at $\eta_{\mathrm{s}}=0$ from hydrodynamic simulations
are plotted as a function of proper time
in the collision of impact parameter $b=3.35$ $(12.1)$ fm 
 in upper (lower) panel of Fig.~\ref{fig:timeevoldens}.
As seen in the figures, thermalized parton density drops rapidly due to the
strong longitudinal  expansion of the system produced in heavy ion collisions.
We expect that jets are likely to loss their energies
in the time span of less than $\tau \sim  2$ fm/$c$.
In order to see when jets lose their energies,
 we plot in Fig.~\ref{fig:jetquenchingrate},
the ratio of the numbers of jets $N(\tau)/N(\tau_0)$
at $p_{\mathrm{T}}=5$ GeV/$c$.
It is found that about 90\% of jet quenching occurs up to $\tau \sim 2$ fm/$c$,
although the QGP phase exists up to $\tau \sim 4$ fm/$c$
and the mixed phase lasts until $\tau \sim 10$ fm/$c$
within the hydro parameters used in this work.
This holds true,
even when we take the total density neglecting the existence of the hadrons
in the mixed phase in Eq.~(\ref{eq:energyloss})
in order to get the maximum energy loss.
Therefore, within our model,
only QGP phase is responsible for the jet energy loss.

%
We showed in Fig.~\ref{fig:dndptpi0_central}
the energy loss $dE/dx=0.2$ GeV/fm gives the same
amount of energy loss as
the energy loss parameter $\epsilon\sigma=0.06$ GeV fm$^2$.
%
Let us estimate a space-time averaged energy loss parameter
for central collision ($b=3.35$ fm).
If we take a space-time average parton density $\bar{\rho}(\tau)$,
 $\bar{\rho}(\tau\leq\text{1.0 fm/}c)=11$ fm$^{-3}$,
 $\bar{\rho}(\tau\leq\text{2.0 fm/}c)=7.7$ fm$^{-3}$,
and  $\bar{\rho}(\tau\leq\text{3.0 fm/}c)=6.0$ fm$^{-3}$,
then the space-time averaged energy loss
yields
$dE/dx(\tau \leq \text{1.0 fm/}c) \sim 0.66$ GeV/fm,
$dE/dx(\tau \leq \text{2.0 fm/}c) \sim 0.46$ GeV/fm,
and $dE/dx(\tau \leq \text{3.0 fm/}c) \sim 0.36$ GeV/fm
 with the best fit  parameter of
$\epsilon\sigma=0.06$ GeV fm$^2$.
Those values are close to that of the
result calculated by the pQCD parton model~\cite{Wang:2001gv},
in which the effect of expansion is not included.
The reason is that the energy loss parameter
can be regarded as an average of roughly
a short time of $\tau \leq 2$ fm/$c$.

In summary,
  we proposed the full 3D hydrodynamical model combined with (mini-) jets,
where jets are explicitly propagated in space-time with
the hydro simulation. In particular,
 this model allows us to study the dynamical
effects on the jet energy loss.
 We estimated the energy loss of partons by the fluid elements
whose temperature is above the critical value $T_c$.
It is found that energy loss of jets occurs in the pure QGP
phase of $\tau < 2$ fm/$c$ and that the contribution of energy loss in the 
mixed phase is negligible under the assumption of the
energy loss formula Eq.~(\ref{eq:energyloss}).
This indicates that suppression of hadronic high $p_{\mathrm{T}}$ spectra
 contain information about the early stage of \textit{partonic} matter.
However, it has been shown that the formula for the energy loss
has non-trivial energy dependence~\cite{Baier,Levai}
and it is found that energy loss is sensitive to
 the critical point~\cite{DP01}.
It is very important to take into account the coherent
 (Landau-Pomeranchuk-Migdal) effect
 in the calculation of radiation spectrum.
A result obtained by using a different energy loss formula
will be presented elsewhere.

A prediction at higher transverse momentum region is
under progress.  Elliptic flow parameter should be also studied
within the model.
It should be studied the charged particle spectra
in order to have the unified understanding of the jet quenching
mechanism in the medium.

In this study, we do not include the energy loss effects
before hydrodynamical evolution.
It might be interesting to study to what extent
partons lose energy before thermalization
using non-equilibrium models~\cite{Nara:2001zu,AR99}.
Because parton density is maximum in this stage,
energy loss by these partons should be important.

\begin{acknowledgments}
We would like to thank  T. Hatsuda,
 J.~Jalilian-Marians,
 and
 A. Dumitru
 for useful comments.
 We would like to thank R. Venugopalan for reading the manuscript.
\end{acknowledgments}


\end{document}